
\catcode`@=11


\font\seventeenrm=cmr17
\font\seventeeni=cmmi12 scaled\magstep2
\font\seventeensy=cmsy10  scaled\magstep3
\font\seventeenex=cmex10  scaled\magstep3
\font\seventeengreek=cmr12 scaled\magstep2
\font\seventeenbf=cmbx12 scaled\magstep2
\font\seventeenib=cmmib10 scaled\magstep3
\font\seventeensyb=cmbsy10 scaled\magstep3
\font\seventeenexb=cmexb10 scaled\magstep3

\font\seventeengreekb=cmbx12 scaled\magstep2

\font\fourteenrm=cmr12   scaled\magstep1
\font\fourteeni=cmmi12   scaled\magstep1
\font\fourteensy=cmsy10  scaled\magstep2
\font\fourteenex=cmex10  scaled\magstep2
\font\fourteengreek=cmr12 scaled\magstep1
\font\fourteenbf=cmbx12  scaled\magstep1
\font\fourteensc=cmcsc10 scaled\magstep2
\font\fourteenib=cmmib10 scaled\magstep2

\font\twelverm=cmr12
\font\twelvei=cmmi12
\font\twelvesy=cmsy10  scaled\magstep1
\font\twelveex=cmex10  scaled\magstep1
\font\twelvebf=cmbx12
\font\twelveit=cmti12
\font\twelvebfit=cmbxti12
\font\twelvesl=cmsl12
\font\twelvebfsl=cmbxsl10 scaled\magstep1
\font\twelvett=cmtt12
\font\twelvebftt=cmbtt10 scaled\magstep1
\font\twelvesc=cmcsc10 scaled\magstep1
\font\twelveib=cmmib10 scaled\magstep1
\font\twelvesyb=cmbsy10 scaled\magstep1
\font\twelveexb=cmexb10 scaled\magstep1
\font\twelvemsam=msam10 scaled\magstep1
\font\twelvebb=msbm10  scaled\magstep1
\font\twelvegoth=eufm10    scaled\magstep1
\font\twelvescript=eusm10  scaled\magstep1
\font\twelvegreek=cmr12
\font\twelvegreekb=cmbx12

\font\tensc=cmcsc10
\font\tenib=cmmib10
\font\tensyb=cmbsy10
\font\tenexb=cmexb10
\font\tenbfsl=cmbxsl10
\font\tenbfit=cmbxti10
\font\tenbftt=cmbtt10
\font\tenmsam=msam10
\font\tenbb=msbm10
\font\tengoth=eufm10
\font\tenscript=eusm10
\font\tengreek=cmr10
\font\tengreekb=cmbx10

\font\ninerm=cmr9
\font\ninei=cmmi9
\font\ninesy=cmsy9

\font\eightrm=cmr8
\font\eighti=cmmi8
\font\eightsy=cmsy8
\font\eightbf=cmbx8
\font\eightsl=cmsl8
\font\eightit=cmti8
\font\eighttt=cmtt8
\font\eightsc=cmcsc8

\font\sevenmsam=msam7
\font\sevenbb=msbm7
\font\sevengoth=eufm7
\font\sevenscript=eusm7
\font\sevengreek=cmr7

\font\fivemsam=msam5


\def\seventeenm@th{%
 \textfont0=\seventeenrm \scriptfont0=\fourteenrm \scriptscriptfont0=\twelverm
 \textfont1=\seventeeni  \scriptfont1=\fourteeni  \scriptscriptfont1=\twelvei
 \textfont2=\seventeensy \scriptfont2=\fourteensy \scriptscriptfont2=\twelvesy
 \textfont3=\seventeenex \scriptfont3=\fourteenex \scriptscriptfont3=\twelveex}

\def\seventeenm@thbf{%
 \textfont0=\seventeenbf  \scriptfont0=\fourteenbf \scriptscriptfont0=\twelverm
 \textfont1=\seventeenib  \scriptfont1=\fourteenib  \scriptscriptfont1=\twelvei
 \textfont2=\seventeensyb \scriptfont2=\fourteensy \scriptscriptfont2=\twelvesy
 \textfont3=\seventeenexb \scriptfont3=\fourteenex \scriptscriptfont3=\twelveex}

\def\fourteenm@th{%
 \textfont0=\fourteenrm  \scriptfont0=\tenrm  \scriptscriptfont0=\sevenrm
 \textfont1=\fourteeni   \scriptfont1=\teni   \scriptscriptfont1=\seveni
 \textfont2=\fourteensy  \scriptfont2=\tensy  \scriptscriptfont2=\sevensy
 \textfont3=\fourteenex  \scriptfont3=\tenex  \scriptscriptfont3=\tenex}

\def\twelvem@th{%
 \textfont0=\twelverm  \scriptfont0=\ninerm  \scriptscriptfont0=\sevenrm
 \textfont1=\twelvei   \scriptfont1=\ninei   \scriptscriptfont1=\seveni
 \textfont2=\twelvesy  \scriptfont2=\ninesy  \scriptscriptfont2=\sevensy
 \textfont3=\twelveex  \scriptfont3=\tenex   \scriptscriptfont3=\tenex}

\def\twelvem@thbf{%
 \textfont0=\twelvebf  \scriptfont0=\tenbf  \scriptscriptfont0=\sevenbf
 \textfont1=\twelveib  \scriptfont1=\tenib  \scriptscriptfont1=\seveni
 \textfont2=\twelvesyb \scriptfont2=\ninesy  \scriptscriptfont2=\sevensy
 \textfont3=\twelveexb \scriptfont3=\tenex   \scriptscriptfont3=\tenex}

\def\tenm@th{%
 \textfont0=\tenrm  \scriptfont0=\sevenrm  \scriptscriptfont0=\fiverm
 \textfont1=\teni   \scriptfont1=\seveni   \scriptscriptfont1=\fivei
 \textfont2=\tensy  \scriptfont2=\sevensy  \scriptscriptfont2=\fivesy
 \textfont3=\tenex  \scriptfont3=\tenex    \scriptscriptfont3=\tenex}

\def\tenm@thbf{%
 \textfont0=\tenbf  \scriptfont0=\sevenrm  \scriptscriptfont0=\fiverm
 \textfont1=\tenib  \scriptfont1=\seveni   \scriptscriptfont1=\fivei
 \textfont2=\tensyb \scriptfont2=\sevensy  \scriptscriptfont2=\fivesy
 \textfont3=\tenexb \scriptfont3=\tenex    \scriptscriptfont3=\tenex}

\def\eightm@th{%
 \textfont0=\eightrm  \scriptfont0=\fiverm  \scriptscriptfont0=\fiverm
 \textfont1=\eighti   \scriptfont1=\fivei   \scriptscriptfont1=\fivei
 \textfont2=\eightsy  \scriptfont2=\fivesy  \scriptscriptfont2=\fivesy
 \textfont3=\tenex    \scriptfont3=\tenex   \scriptscriptfont3=\tenex}

\def\sevenm@th{%
 \textfont0=\sevenrm  \scriptfont0=\fiverm  \scriptscriptfont0=\fiverm
 \textfont1=\seveni   \scriptfont1=\fivei   \scriptscriptfont1=\fivei
 \textfont2=\sevensy  \scriptfont2=\fivesy  \scriptscriptfont2=\fivesy
 \textfont3=\tenex    \scriptfont3=\tenex   \scriptscriptfont3=\tenex}




\newfam\greekf@m
\newfam\msamf@m
\newfam\bbf@m
\newfam\gothf@m
\newfam\scriptf@m


\def\Gamma{{\fam\greekf@m\mathchar"7800}}
\def\Delta{{\fam\greekf@m\mathchar"7801}}
\def\Theta{{\fam\greekf@m\mathchar"7802}}
\def\Lambda{{\fam\greekf@m\mathchar"7803}}
\def\Xi{{\fam\greekf@m\mathchar"7804}}
\def\Pi{{\fam\greekf@m\mathchar"7805}}
\def\Sigma{{\fam\greekf@m\mathchar"7806}}
\def\Upsilon{{\fam\greekf@m\mathchar"7807}}
\def\Phi{{\fam\greekf@m\mathchar"7808}}
\def\Psi{{\fam\greekf@m\mathchar"7809}}
\def\Omega{{\fam\greekf@m\mathchar"780A}}


\def\seventeenpoint{%
  \def\lf{%
     \textfont\greekf@m=\seventeengreek
     \scriptfont\greekf@m=\fourteengreek
     \def\rm{\seventeenm@th\fam0\seventeenrm}%
     \def\it{\sevenit}%
     \def\sl{\sevensl}%
     \def\tt{\seventt}%
     \def\sc{\sevensc}%
     \rm}%
  \def\bf{%
     \textfont\greekf@m=\seventeengreekb
     \scriptfont\greekf@m=\fourteengreek
     \def\rm{\seventeenm@thbf\fam0\seventeenbf}%
     \rm}%
  \normalbaselineskip=20pt\normalbaselines
  \lf}

\def\fourteenpoint{%
  \def\lf{\def\rm{\fourteenm@th\fam0\fourteenrm}\rm}%
  \def\bf{\def\rm{\fourteenbf}\fam0\rm}%
  \def\sc{\fourteensc}%
  \normalbaselineskip=17pt\normalbaselines
  \lf}

\def\twelvepoint{%
  \textfont\msamf@m=\twelvemsam
  \scriptfont\msamf@m=\tenmsam
  \textfont\bbf@m=\twelvebb
  \scriptfont\bbf@m=\tenbb
  \scriptscriptfont\bbf@m=\sevenbb
  \textfont\gothf@m=\twelvegoth
  \scriptfont\gothf@m=\tengoth
  \scriptscriptfont\gothf@m=\sevengoth
  \textfont\scriptf@m=\twelvescript
  \scriptfont\scriptf@m=\tenscript
  \scriptscriptfont\scriptf@m=\sevenscript
  \def\lf{%
      \textfont\greekf@m=\twelvegreek
      \scriptfont\greekf@m=\tengreek
      \scriptscriptfont\greekf@m=\sevengreek
      \def\rm{\twelvem@th\fam0\twelverm}%
      \def\it{\twelveit}%
      \def\sl{\twelvesl}%
      \def\tt{\twelvett}%
      \def\sc{\twelvesc}%
      \rm}%
  \def\bf{%
      \textfont\greekf@m=\twelvegreekb
      \scriptfont\greekf@m=\tengreekb
      \def\rm{\twelvem@thbf\fam0\twelvebf}%
      \def\it{\twelvebfit}%
      \def\sl{\twelvebfsl}%
      \def\tt{\twelvebftt}%
      \def\sc{\twelvebfsc}%
      \rm}%
  \def\msam{\fam\msamf@m\twelvemsam}%
  \def\bb{\fam\bbf@m\twelvebb}%
  \def\goth{\fam\gothf@m\twelvegoth}%
  \def\script{\fam\scriptf@m\twelvescript}%
  \normalbaselineskip=14pt\normalbaselines
  \lf}

\def\tenpoint{%
  \textfont\msamf@m=\tenmsam
  \scriptfont\msamf@m=\sevenmsam
  \scriptscriptfont\msamf@m=\fivemsam
  \textfont\bbf@m=\tenbb
  \scriptfont\bbf@m=\sevenbb
  \textfont\gothf@m=\tengoth
  \scriptfont\gothf@m=\sevengoth
  \textfont\scriptf@m=\tenscript
  \scriptfont\scriptf@m=\sevenscript
  \def\lf{%
      \textfont\greekf@m=\tengreek
      \scriptfont\greekf@m=\sevengreek
      \def\rm{\tenm@th\fam0\tenrm}%
      \def\it{\tenit}%
      \def\sl{\tensl}%
      \def\tt{\fam\ttfam\tentt}%
      \def\sc{\tensc}%
      \rm}%
  \def\bf{%
      \textfont\greekf@m=\tengreekb
      \def\rm{\tenm@thbf\fam0\tenbf}%
      \def\it{\tenbfit}%
      \def\sl{\tenbfsl}%
      \def\tt{\fam\ttfam\tenbftt}%
      \def\sc{\tenbfsc}%
      \rm}%
  \def\msam{\fam\msamf@m\tenmsam}%
  \def\bb{\fam\bbf@m\tenbb}%
  \def\goth{\fam\gothf@m\tengoth}%
  \def\script{\fam\scriptf@m\tenscript}%
  \normalbaselineskip=12pt\normalbaselines
  \lf}

\def\eightpoint{%
  \def\lf{\def\rm{\eightm@th\fam0\eightrm}\rm}%
  \def\bf{\def\rm{\eightbf}\rm}%
  \def\it{\eightit}%
  \def\sl{\eightsl}%
  \def\tt{\eighttt}%
  \def\sc{\eightsc}%
  \normalbaselineskip=10pt\normalbaselines
  \setbox\strutbox=\hbox{\vrule height7pt depth3pt width0pt}%
  \lf}

\def\sevenpoint{%
  \def\lf{\def\rm{\sevenm@th\fam0\sevenrm}\rm}%
  \def\bf{\def\rm{\sevenbf}\rm}%
  \def\it{\sevenit}%
  \def\sl{\sevensl}%
  \def\tt{\eighttt}
  \normalbaselineskip=8pt\normalbaselines
  \lf}

\tenpoint

\def\mathbf#1{\hbox{\bf$#1$}}


\newif\ifreffile          \reffilefalse
\newread\reffile
\newread\pagenosfile
\newwrite\reffile

\outer\def\openreffile{%
   \re@dreffile
   \immediate\write16{Writing references to \jobname.ref}%
   \immediate\openout\reffile=\jobname.ref
   \reffiletrue}

\def\re@dreffile{%
     \openin\reffile=\jobname.ref
     \ifeof\reffile
        \closein\reffile
        \immediate\write16{*** Not found \jobname.ref ***}%
     \else\closein\reffile
        \immediate\write16{Reading references from \jobname.ref}%
     \let\prefixforref=\relax
     \catcode`@=11
     \input\jobname.ref
     \catcode`@=12
     \fi}

\newread\xreffile
\outer\def\readref#1{%
   \openin\xreffile=#1.ref
   \ifeof\xreffile
       \closein\xreffile
       \immediate\write16{*** Not found #1.ref ***}%
   \else\closein\xreffile
        \immediate\write16{Reading references from #1.ref}%
        \def\prefixforref{\xchapno.}%
        \catcode`@=11  
        \input#1.ref
        \catcode`@=12
        \let\prefixforref=\relax
   \fi
   }



\def\landscape{{\let\tmpsize=\hsize
                \global\let\hsize=\vsize
		\global\let\vsize=\tmpsize}}

\def\t@define{$\spadesuit$}

\def\getp@geno{\global\pageno=\pagenos\chapterno}

\newif\ifshowref    \showreffalse
\let\innerl@bel=\relax
\def\label#1{\def\innerl@bel{#1}\par}  


\def\frenchmonth{\ifcase\month\or janvier\or f\'evrier\or mars\or avril%
       \or mai\or juin\or juillet\or ao\^ut%
       \or septembre\or octobre\or novembre\or d\'ecembre\fi}
\def\englishmonth{\ifcase\month\or January\or February\or March\or April%
       \or May\or June\or July\or August%
       \or September\or October\or November\or December\fi}
\def\frenchdate{\the\day\ \frenchmonth\ \the\year}
\def\englishdate{\englishmonth\ \the\day, \the\year}
\let\date=\frenchdate
\newif\ifshowdate       \showdatetrue
\def\author{F.~Arnault}
\newif\iffirstpage      \firstpagetrue
\newtoks\firstheadline
\newtoks\rightheadline
\newtoks\leftheadline
\newtoks\firstfootline
\newtoks\otherfootline

\headline={%
   \iffirstpage\the\firstheadline
   \else\ifodd\pageno\the\rightheadline
        \else\the\leftheadline\fi
   \fi}
\firstheadline={\hfil}
\rightheadline={\tenpoint\sl\hfil\firstmark\hfil}
\leftheadline={\tenpoint\sl\hfil\firstmark\hfil}

\footline={%
   \iffirstpage\the\firstfootline\global\firstpagefalse
   \else\the\otherfootline
   \fi}
\firstfootline={\ifshowdate \hfil\sevenrm\author\quad---\quad\date\hfil 
                \else\footfolio \fi}
\otherfootline={\footfolio}
\def\footfolio{\hfil\tenrm --- \folio\ ---\hfil}


\newskip\titleskipamount
\titleskipamount=24pt plus4pt minus4pt  
\def\chapterno{\t@define}
\def\ch@pterm@rk{\relax}

\outer\def\chapter#1#2 #3\par{
   \global\edef\chapterno{#1}%
   \ifnum#1>-1
       \global\edef\ch@pterm@rk{\chapterno.}%
   \fi
   \re@dp@genos
   \getp@geno
   \ifreffile
      \immediate\write\reffile{\def\string\xchapno{#1}}%
      \immediate\write\reffile{\def\string\chap#2{#1}}%
   \fi
   \message{#2 -> chapitre #1 : #3. }%
   {\def\cr{\relax}\mark{#3}}%
   \leftheadline={\tenpoint\sl\hfil\ch@pterm@rk{\def\cr{\relax} #3}\hfil}
   \ifnum#1>-1
     \leftline{\twelvepoint\bf Chapitre #1}%
     \bigskip
   \fi
   {\twentyonepointb\halign{\centerline{##}\cr#3\crcr}}%
   \ifreffile
      \write\reffile{\string\t@cchapter{\folio}{#1}{#3}}%
   \fi
   \bigskip\bigskip\bigskip
   }

\outer\def\title#1\par{%
   {\def\cr{\relax}\mark{#1}}%
   {\seventeenpoint\bf\halign{\centerline{##}\cr#1\crcr}}%
   \ifreffile
      \immediate\write\reffile{\def\string\xchapno{\relax}}%
      \immediate\write\reffile{\string\t@ctitle{#1}}%
   \fi
   \vskip\titleskipamount}

\outer\def\subtitle#1\par{%
   \vskip-\titleskipamount
   {\fourteenpoint\halign{\centerline{##}\cr#1\crcr}}%
   \vskip\titleskipamount}

\parskip=2pt plus 1pt minus 0.5pt   


\newdimen\abstractmarginswidth
\abstractmarginswidth=30pt
\long\def\abstract#1.#2\endabstract{%
  \begingroup
  \parindent=\abstractmarginswidth
  \narrower\noindent\eightpoint
  {\bf#1.\enspace}%
  #2\par
  \endgroup}
\def\endabstract{%
 \errhelp={I have encountered a end-of-abstract mark, outside of any abstract}%
 \errmessage{\string\endabstract\space must match an \string\abstract}}


\newif\ifromansection \romansectionfalse
\def\stringsectionnumber#1{\ifromansection\uppercase\expandafter{\romannumeral#1}\else\the#1\fi}

\newcount\sectionnumber
\outer\def\section#1 \par{
    \global\advance\sectionnumber by1
    \global\statnumber=0
    \ifdim\lastskip<\bigskipamount \removelastskip\bigskip \fi
    \penalty-400
    \ifx\innerl@bel\relax\relax
    \else
      \expandafter\xdef\csname sec\innerl@bel\endcsname{\the\sectionnumber}%
      \message{\innerl@bel -> section \the\sectionnumber, }%
      \ifreffile
         \immediate\write\reffile{\edef\string\sec\innerl@bel{%
                          \string\prefixforref\stringsectionnumber{\sectionnumber}}}%
      \fi
    \fi
    \leftline{%
          \ifshowref
            \ifx\innerl@bel\relax\relax  \else\llap{\fiverm\innerl@bel\ }\fi
          \fi
          \twelvepoint\bf\stringsectionnumber{\sectionnumber}. #1}%
    \mark{\ch@pterm@rk\stringsectionnumber{\sectionnumber}. #1}%
    \ifreffile
       \write\reffile{\string\t@csecti@n{\folio}%
                                        \string{#1\string}}%
    \fi
    \nobreak\medskip\noindent
    \let\innerl@bel=\relax
    }

\outer\def\subsection#1 \par{%
    \ifhmode  
       \immediate\write-1{%
             I guess that the subsection just follows
             section \the\sectionnumber's title.}%
    \else
        \ifdim\lastskip<\bigskipamount \removelastskip\bigskip \fi
        \penalty-300 
    \fi
    \leftline{\twelvepoint#1}%
    \nobreak\medskip\noindent
    }

\outer\def\subsubsection#1 \par{%
    \ifhmode  
       \immediate\write-1{%
             I guess that the subsubsection just follows a section title.}%
    \else
        \ifdim\lastskip<\bigskipamount \removelastskip\bigskip \fi
        \penalty-250 
    \fi
    \leftline{#1}%
    \nobreak\smallskip\noindent
    }


\newcount\statnumber

\def\innerstat#1#2{%
    \global\advance\statnumber by1
    \ifx\innerl@bel\relax\relax
    \else
       \expandafter\xdef\csname ref\innerl@bel\endcsname
          {\stringsectionnumber{\sectionnumber}.\the\statnumber}%
       \message{\innerl@bel -> \csname ref\innerl@bel\endcsname, }%
       \ifreffile\immediate\write\reffile{%
             \edef\string\ref\innerl@bel{%
             \string\prefixforref\stringsectionnumber{\sectionnumber}.\the\statnumber}}%
       \fi
    \fi
    \medbreak\noindent
    \ifshowref
       \ifx\innerl@bel\relax\relax \else\llap{\fiverm\innerl@bel\ }\fi
    \fi
    \alphaenumreset
    {\bf\stringsectionnumber\sectionnumber.\the\statnumber. --- #1.\enspace}%
    {#2\par}%
    \let\innerl@bel=\relax
    \ifdim\lastskip<\medskipamount \removelastskip\penalty55\medskip\fi
    }

\outer\def\stat #1. #2\par{\innerstat{#1}{\sl#2}}
\outer\def\statrm #1. #2\par{\innerstat{#1}{#2}} 
\outer\def\remark #1. #2\par{%
    \smallbreak\noindent
    {\bf#1.\enspace}%
    {#2\par}%
    \ifdim\lastskip<\medskipamount \removelastskip\penalty55\medskip\fi
    }

\newcount\eqnumber
\def\eqdef#1{%
    \global\advance\eqnumber by1
    \expandafter\xdef\csname eq#1\endcsname{\the\eqnumber}%
    \message{#1 -> (\csname eq#1\endcsname), }%
    \ifreffile\immediate\write\reffile{%
          \edef\string\eq#1{\string\prefixforref\the\eqnumber}}%
    \fi
    \eqno\hbox{%
       \lf(\the\eqnumber)%
       \ifshowref\rlap{\fiverm\ #1}\fi}}

\def\eqaligndef#1{%
    \global\advance\eqnumber by1
    \expandafter\xdef\csname eq#1\endcsname{\the\eqnumber}%
    \message{#1 -> (\csname eq#1\endcsname), }%
    \ifreffile\immediate\write\reffile{\def\string\eq#1{\the\eqnumber}}%
    \fi
    \lf(\the\eqnumber)%
    \ifshowref\rlap{\fiverm\ #1}\fi}

\newcount\enumnumber

\def\enum{%
   \advance\enumnumber by1
   \par\noindent(\the\enumnumber) }

\def\alphaenumreset{\enumnumber="60 }
\def\alphaenum{%
  \advance\enumnumber by1
  \par\noindent(\char\enumnumber) }
\def\rawalphaenum{%
  \advance\enumnumber by1
  (\char\enumnumber) }


\newbox\figb@x
\newcount\fignumber
\newdimen\figb@xwidth

\def\figure#1. #2#3{%
  \global\advance\fignumber by1
  \ifx\innerl@bel\relax\relax
  \else
     \expandafter\xdef\csname fig\innerl@bel\endcsname{\the\fignumber}%
     \message{\innerl@bel -> #1 \csname fig\innerl@bel\endcsname, }%
     \ifreffile\immediate\write\reffile{%
         \edef\string\fig\innerl@bel{%
         \string\prefixforref\the\fignumber}}%
     \fi
  \fi
  \setbox\figb@x=\vbox{#3}%
  \figb@xwidth=\wd\figb@x
  \setbox\figb@x=\vbox{%
       \box\figb@x
       \smallskip
       \hbox to\figb@xwidth{%
          \ifshowref
             \ifx\innerl@bel\relax\relax \else\llap{\fiverm\innerl@bel\ }\fi
          \fi
          \hss{\bf #1\ \the\fignumber.} #2\hss}}%
  $$%
    \box\figb@x
  $$%
  \let\innerl@bel=\relax}


\newbox\bibb@x
\newcount\bibchosen
\newcount\bibnumber
\outer\def\showbib{\unvbox\bibb@x}

\def\bibdef#1#2#3#4{%
    \global\advance\bibnumber by1
    \expandafter\xdef\csname bib#1\endcsname{\the\bibnumber}%
    \global\setbox\bibb@x=\vbox{%
        \unvbox\bibb@x
        \hang\textindent{[\the\bibnumber]}%
        \strut{\sc#2 :\ \it#3.\/\ \ \rm#4.}%
        \par\medskip}}

\def\bibunlo@ded{\t@define}

\def\bibchoose#1{%
    \global\advance\bibchosen by1
    \expandafter\def\csname bib#1\endcsname{\t@define}}

\let\ifbibchooseall=\iffalse
\def\bibchooseall{\let\ifbibchooseall=\iftrue}

\outer\def\checkbib{\ifbibchooseall\relax
         \else\ifnum\bibchosen=\bibnumber\relax
              \else\immediate\write16{Warning : Some bibliographic 
                                      references remain undefined.}%
              \fi
         \fi}

\outer\def\bibitem#1#2#3#4{%
   \ifbibchooseall\bibdef{#1}{#2}{#3}{#4}%
   \else
       \expandafter\ifx\csname bib#1\endcsname\relax \relax %
       \else
           \expandafter\ifx\csname bib#1\endcsname\bibunlo@ded
               \bibdef{#1}{#2}{#3}{#4}%
           \else\immediate\write16{Warning : attempt to redefine
                                   bibliographic reference #1.}%
           \fi
       \fi
   \fi}



\def\chapref#1{%
    \expandafter\ifx\csname chap#1\endcsname\relax \t@define
          \immediate\write16{ *** Undefined chapter reference : #1 ***}%
    \else\csname chap#1\endcsname
    \fi}

\def\secref#1{%
    \expandafter\ifx\csname sec#1\endcsname\relax \t@define
          \immediate\write16{ *** Undefined section reference : #1 ***}%
    \else\csname sec#1\endcsname
    \fi}

\def\ref#1{%
    \expandafter\ifx\csname ref#1\endcsname\relax \t@define
          \immediate\write16{ *** Undefined statement reference : #1 ***}%
    \else\csname ref#1\endcsname
    \fi}

\def\eqref#1{%
    \expandafter\ifx\csname eq#1\endcsname\relax \t@define
          \immediate\write16{ *** Undefined equation reference : #1 ***}%
    \else(\csname eq#1\endcsname)%
    \fi}

\def\figref#1{%
    \expandafter\ifx\csname fig#1\endcsname\relax \t@define
          \immediate\write16{ *** Undefined figure reference : #1 ***}%
    \else\csname fig#1\endcsname
    \fi}

\def\bib#1{%
   \expandafter\ifx\csname bib#1\endcsname\relax
      [\t@define]%
      \immediate\write16{ *** Unknown bibliographic reference : #1 ***}%
   \else
      \expandafter\ifx\csname bib#1\endcsname\bibunlo@ded
          \immediate\write16{ *** Unloaded bibliographic reference : #1 ***}%
      \fi
      [\csname bib#1\endcsname]%
   \fi}


\newif\iftoc   \tocfalse
\newbox\t@cb@x
\newcount\t@csecn@
\newif\ifnarrowtoc  \narrowtoctrue

\def\t@cchapter#1#2#3{
  \iftoc{%
     \global\narrowtocfalse
     \def\cr{\relax}%
     \advance\hsize by -\pagecolwidth
     \global\t@csecn@=0
     \parindent=0pt
     \global\setbox\t@cb@x=\vbox{%
         \unvbox\t@cb@x
         \bigskip\bigskip\goodbreak
         \ifnum#2>-1
           \centerline{\bf Chapitre #2}%
           \smallskip\nobreak
         \fi
         \strut\blacksquare\ \bf#3\page{#1}\kern-\pagecolwidth}%
         \smallskip\nobreak}%
  \else\relax
  \fi}

\def\t@ctitle#1{%
  \iftoc{%
     \global\narrowtoctrue
     \parindent=\abstractmarginswidth
     \narrower
     \def\cr{\relax}%
     \advance\hsize by -\pagecolwidth
     \global\t@csecn@=0
     \parindent=0pt
     \global\setbox\t@cb@x=\vbox{%
         \unvbox\t@cb@x
         \medskip
         \strut\blacksquare\ \bf#1}%
         \medskip\nobreak}%
  \else\relax
  \fi}

\def\t@csecti@n#1#2{%
  \iftoc{%
     \advance\hsize by -\pagecolwidth
     \global\advance\t@csecn@ by1
     \parindent=0pt
     \global\setbox\t@cb@x=\vbox{%
         \unvbox\t@cb@x
             \strut
             \ifnarrowtoc\kern\abstractmarginswidth\fi
             \the\t@csecn@. --- #2\page{#1}%
             \ifnarrowtoc\kern\abstractmarginswidth\fi
             \kern-\pagecolwidth}%
          \smallskip}%
  \else\relax
  \fi}

\def\showtoc{%
  \iftoc\unvbox\t@cb@x
  \else
    \immediate\write16{*** TOC is void because you hav'nt said
                       \string\toctrue. ***}%
  \fi}


\def\pagenos#1{1\write16{Using default pageno : 1}}

\def\re@dp@genos{%
     \openin\pagenosfile=pagenos.tex
     \ifeof\pagenosfile\closein\pagenosfile
     \else\closein\pagenosfile
        \immediate\write16{Reading pagenos.tex}%
     \input pagenos.tex
     \fi}

\everyjob={%
     \immediate\write16{Format Light AzTeX (F. Arnault), version \fmtversion.}
     \setbox\t@cb@x=\vbox{}%
     \setbox\bibb@x=\vbox{}%
     \tracingstats=1}



{\obeyspaces\global\let =\ }  
{\def\tmpminus{-\relax}\catcode`-=\active \xdef-{\tmpminus}}
{\def\tmpgreater{>\relax}\catcode`>=\active \xdef>{\tmpgreater}}
{\def\tmplessthan{<\relax}\catcode`<=\active \xdef<{\tmplessthan}}

\def\unc@tcodespeci@ls{\def\do##1{\catcode`##1=12}\dospecials}

\outer\def\verbatim{\par\begingroup\setupverb@tim\doverbatim}

\def\setupverb@tim{%
  \parindent=0mm
  \def\par{\leavevmode\endgraf}
  \obeylines\unc@tcodespeci@ls\catcode`-=\active\catcode`>=\active
       \catcode`<=\active\obeyspaces\tt}

{\catcode`\|=0 \catcode`\\=12  
 |obeylines|gdef|doverbatim^^M#1\endverbatim
 {|medskip|hrule|nobreak#1|smallskip|nobreak|hrule|medskip|endgroup|noindent}}

\def\endverbatim{%
  \errhelp={I have encountered an end-verbatim mark whereas I was not in
           verbatim mode}%
  \errmessage{\string\endverbatim\space must match a \string\verbati m}}

{\catcode`\|=\active\obeylines%
\gdef\|{\begingroup\catcode`\|=\active\setupverb@tim\let^^M=\ \let|=\endgroup}}

\newcount\l@st@ngl@ne
\outer\def\listing#1#2\par{%
   \par\begingroup\setupverb@tim
   \everypar{\advance\l@st@ngl@ne by1\llap{\sevenrm\the\l@st@ngl@ne\quad}}%
   \medskip
   \line{\leaders\hrule\hfil\quad\sl#1\quad\leaders\hrule\hfil}\nobreak
   \input#2
   \nobreak\smallskip\nobreak\hrule\medskip
   \endgroup}



\newif\ifshowhidden  \showhiddenfalse

\def\endhidden{\ifshowhidden\relax\else
  \errhelp={I have encountered a end-of-hidden-part mark, 
            without matching hidden-part mark.}%
  \errmessage{\string\endhidden\space must match a \string\hidde n}\fi}

\outer\def\hidden{\ifshowhidden\relax
                  \else\par\begingroup\s@tuphidd@n\dohidden\fi}

\def\s@tuphidd@n{\obeylines\unc@tcodespeci@ls}  

{\catcode`\|=0 \catcode`\\=12  
 |obeylines|gdef|dohidden#1\endhidden{|endgroup|fi}}  


\font\initialfont=cmbx12 scaled\magstep4
\newbox\initialb@x
\def\initial#1{%
   \setbox\initialb@x=\hbox{\initialfont#1\hskip2pt}%
   \hang\hangafter=-2\hangindent=\wd\initialb@x
   \setbox\initialb@x=\hbox{\hskip-\wd\initialb@x\box\initialb@x}%
   \noindent
   \smash{\lower12pt\box\initialb@x}%
   }
 
\catcode`@=12


\newdimen\pagecolwidth  \pagecolwidth=8mm
\def\page#1{%
       \quad
       \leaders\hrule height0.1pt\hfill
       \hbox to\pagecolwidth{\hfill#1}}
       
\def\qed{\kern 4pt\penalty500\null\hfill\square\par}

\def\square{%
   \hbox{%
      \vrule
      \vbox to 6pt{\hrule width 4pt\vfill\hrule}%
      \vrule}}

\def\blacksquare{%
  \vbox{%
    \hbox{%
      \kern1pt
      \vrule height5pt depth 0pt width 5pt
      \kern1pt}%
    \kern0.5pt}}

\def\pdem#1-{\vskip-4pt\noindent{\sc#1 }--- \alphaenumreset}
\def\proof{\vskip-3pt\noindent{\sc Proof --- }\alphaenumreset}

\def\C{{\bb C}}


\def\hbar{{\mathord{\bb\mathchar"717D}}}  

\def\leq{\mathrel{\msam\mathchar"7136}}
\def\geq{\mathrel{\msam\mathchar"713E}}


\def\ceil#1{{\left\lceil#1\right\rceil}}
 
\def\ket#1{{\mathopen\vert#1\rangle}}

\def\braket#1#2{{\langle#1\vert#2\rangle}}

\def\angle#1{{\langle#1\rangle}}

\def\abs#1{\mathopen\vert#1\mathclose\vert}  
\def\Abs#1{{\left|#1\right|}}

\def\assign{\mathrel{:=}}


\def\\{\par}  

\def\qbox#1{\quad\hbox{#1}\quad}
\def\qqbox#1{\qquad\hbox{#1}\qquad}






\def\prob{\mathop{\rm prob}\nolimits}

\def\SU{\mathop{\rm SU}\nolimits}

\def\fmtversion{2.6.6}

\showdatefalse

\title Random generation with the spin of a qutrit

\vskip-4mm
\centerline{Fran\c cois ARNAULT (arnault@unilim.fr) \& Don ANOMAN (don.anoman@unilim.fr)}
\smallskip
\centerline{Universit\'e de Limoges, XLIM/DMI, 123 avenue Albert Thomas, 87060 Limoges, France}
\bigskip

\bibchoose{BCKS}
\bibchoose{BABOE}
\bibchoose{BCKKK}
\bibchoose{CHSH}
\bibchoose{ColbeckKent}
\bibchoose{KKCZO}  
\bibchoose{KCBS}
\bibchoose{KJPWF}  
\bibchoose{KKLM}
\bibchoose{LWLBRGW}
\bibchoose{LBLLBMZB}
\bibchoose{MayersYao}
\bibchoose{PAM+}
\bibchoose{Tsirelson0}
\bibchoose{UZZWSDDK}
\bibchoose{Wootters}

\bibitem{BCKS}
 {S.~Binicio\u{g}lu, M. Ali Can, A.A.~Klyachko, A.S.~Shumovsky}
 {Entanglement of single spin-1 object: an example of ubiquitous entanglement}
 {Foundations of Physics~37, 1253 (2007)} 

\bibitem{BABOE}
 {I.~Bregman, D.~Aharonov, M.~Ben-Or, H.S.~Eisenberg}
 {Simple and secure quantum key distribution with biphotons}
 {Physical Review A~77, 050301(R) (2008)}
 
\bibitem{BCKKK}
 {A.V.~Burlakov, M.V.~Chekhova, O.A.~Karabutova, D.N.~Klyshko, S.P.~Kulik}
 {Polarization state of biphotons: quantum ternary logic}
 {Physical Review A~60, R4209 (1999)}

\bibitem{CHSH}
 {J.F.~Clauser, M.A.~Horne, A.~Shimony, R.A.~Holt}
 {Proposed experiment to test local hidden variables theories}
 {Physical Review Letters~23, 880 (1969)}

\bibitem{ColbeckKent}
 {R.~Colbeck, A.~Kent}
 {Private randomness expansion with untrusted devices}
 {Journal of Physics~A: Math.\ Theor.~44, 095305 (2011)}

\bibitem{KKCZO}
 {D.~Kaszlikowski, L.C.~Kwek, J.L.~Chen, M.~\^Zukowski, C.H.~Oh}
 {Clauser-Horne inequality for three-state systems}
 {Physical Review A~65, 032118 (2002)}

\bibitem{KCBS}
 {A.A.~Klyachko, M.A.~Can, S.~Binicio\u glu, A.S.~Shumovsky}
 {Simple test for hidden variables in spin-1 systems}
 {Physical Review Letters~101, 020403 (2008)}

\bibitem{KJPWF}
 {A.~Kulikov, M.~Jerger, A. Poto\v cnik, A.~Wallraff, A.~Fedorov}
 {Realization of a quantum random generator certified with the Kochen-Specker theorem}
 {Physical Review Letters~119, 240501 (2017)}

\bibitem{KKLM}
 {P.~Kurzy\'nski, A.~Kolodziejski, W.~Laskowski, M.~Markiewicz}
 {Three-dimensional visualization of a qutrit}
 {Physical Review A~93, 062126 (2016)}

\bibitem{LWLBRGW}
 {B.P.~Lanyon, T.J.~Weinhold, N.K.~Langford, J.L.~O'Brien, K.J.~Resch, A.~Gilchrist, A.G.~White}
 {Manipulating biphotonic qutrits}
 {Physical Review Letters~100, 060504 (2008)}

\bibitem{LBLLBMZB}
 {T.~Lunghi, J.B.~Brask, C.~Ci Wen Lim, Q.~Lavigne, J.~Bowles, A.~Martin, H.~Zbinden, N.~Brunner}
 {Self-testing quantum random number generator}
 {Physical Review Letters~114, 150501 (2015)}

\bibitem{MayersYao}
 {D.~Mayers, A.C.~Yao}
 {Quantum cryptography with imperfect apparatus}
 {IEEE Symposium on Foundations of Computer Science (FOCS'98)}

\bibitem{PAM+}
 {S.~Pironio, A.~Ac\'\i n, M.~Massar, A.~Boyer de la Giroday, D.N.~Matsukevich, P.~Maunz,
  S.~Omeschenk, D.~Hayes, L.~Luo, T.A.~Manning, C.~Monroe}
 {Random numbers certified by Bell's Theorem}
 {Nature Letters~464, 1021, doi:10.1038/nature09008 (also: arXiv:0911.3427 [quant-ph]) (2010)}

\bibitem{Tsirelson0}
 {B.S.~Cirel'son}
 {Quantum generalizations of Bell's inequality}
 {Letters in Mathematical Physics~4, 93--100 (1980)}

\bibitem{UZZWSDDK} 
 {M.~Um, X.~Zhang, J.~Zhang, Y.~Wang, Y.~Shen, D.-L.~Deng, L.-M.~Duan, K.~Kim}
 {Experimental certification of random numbers via quantum contextuality}
 {Nature Scientific Reports~3, 1627 (2013)}

\bibitem{Wootters}
 {W.K.~Wootters}
 {Entanglement of formation of an arbitrary state of two qubits}
 {Physical Review Letters~80, 2245 (1998)}

\checkbib

\def\SU{\mathop{\rm SU}}
\def\CHSH{\mathop{\rm CHSH}}

\abstract Abstract.
In this paper, we consider the use of a single qutrit for random generation.  This is possible because single qutrits exhibit contextuality
features~\bib{KCBS}.  Moreover this has yet been realized as reported in~\bib{KJPWF} and~\bib{UZZWSDDK}.  Here, we aim to optimize the entropy of the generated
sequence.  To do this, we do not rely on the KCBS inequality as done in~\bib{UZZWSDDK}, but instead on the use of a specific state and a check for fidelity.  By
the way, we show that this check can be considered as a variant of the CHSH inequality applied to pairs of photons or spin-1/2 particles (qutrits are often
realized as a pair of indistinguishable qubits).  The physical realisation of this random generator should be eased by the fact it needs only to implement spin
operations and measurement, not general $\SU(3)$ qutrit manipulations.
\endabstract
\bigskip

Random numbers are essential for cryptographic applications.  They also have many other uses as simulations of physical processes.
However, secure random numbers (which no adversary can predict) are notoriously hard to produce.  Moreover the deterministic nature of classical
physics forbids the existence of truly random numbers.  Hence classical random generation methods produces numbers which are not truly random but
indistinguishable from truly random (under computational assumptions).

But quantum physics describes processes which are genuinely random and which can be used to produce true random numbers.  Moreover, works as
\bib{ColbeckKent}, \bib{LBLLBMZB} and~\bib{PAM+} show that the amount of randomness of sequences generated by some processes can be lower bounded.  Hence
not only true random generators do exist but they can be certified, even if the devices used for production are untrusted (device independence~\bib{MayersYao}).

In this paper, we consider the use of a single qutrit for random generation.  This is possible because single qutrits exhibit contextuality
features~\bib{KCBS}.  Moreover this has yet been realized as reported in~\bib{KJPWF} and~\bib{UZZWSDDK}.  Here, we aim to optimize the entropy of the generated
sequence.  To do this, we do not rely on the KCBS inequality as done in~\bib{UZZWSDDK}, but instead on the use of a specific state and a check for fidelity.  By
the way, we show that this check can be considered as a variant of the CHSH inequality applied to pairs of photons or spin-1/2 particles (qutrits are often
realized as a pair of indistinguishable qubits).  The physical realisation of this random generator should be eased by the fact it needs only to implement spin
operations and measurement, not general $\SU(3)$ qutrit manipulations.

The organisation of the paper is as follows.  In Section~1, we review what we need about qutrits and their implementation as biphotons.  In Section~2, we
disgress on the Bell inequality CHSH which, by using symmetrisation, we apply to a pair of indistinguishable qubits.  Section~3 considers general spin
measurements, and testing of a particular state.  Section~4 exposes our random generator.

\section Qutrits

Besides qubits which are most often considered for quantum computation, are qutrits which can be represented as state vectors in~$\C^3$.
Qutrits have a lot of useful features for storing quantum information.  They are known to be more robust than qubits against decoherence.  They lead also to
Bell inequalities~\bib{KKCZO} which are more resistant to noise.  Even more importantly for the purpose of the present paper, a single
qutrit exhibits contextuality features, as shown for example in~\bib{KCBS}.

\subsection Biphotonic qutrits

A spin-1 particle carry qutrit quantum information.  But often in practice, qutrits are physically realized as pairs of indistinguishable photons (biphotons)
or by two indistinguishable spin-$1/2$ particles.  The state of the two particles belongs in the space $\C^2\otimes\C^2$ of states of two qubits.  In this
space, the {\sl singlet} state ${1\over\sqrt2}\big(\ket{01}-\ket{10}\big)$ negated if left and right qubits are permuted.  The orthogonal space
of the singlet is spanned by the three states
$$
  \ket{00},
  \qquad
  {1\over\sqrt2} \big(\ket{01} + \ket{10}\big),
  \qquad
  \ket{11}. 
  \eqdef{symbasis}
$$
and is made of symmetric states (invariant when left and right qubits are permuted).  This space of symmetric states is the space of states of two
indistinguishable qubits.  The basis states of~\eqref{symbasis} are sometimes written instead
$$
  \ket+,
  \qquad
  \ket0,
  \qquad
  \ket-
$$
and we adopt this notation.

Note that biphoton qutrits have yet been used to improve security and efficiency of quantum key distribution~\bib{BABOE}.
They also have been used for random generation~\bib{KJPWF}\bib{UZZWSDDK}.

\subsection Qutrit spin

In principle, general action over one qutrit can be described by any operator in the~$\SU(3)$ group, which is a real manifold of dimension~8.  However, spin
operations, which involve the group $\SU(2)$ --- a manifold of dimension~3 --- are much more common and easy to implement
(However, some works such as~\bib{BCKKK} and~\bib{LWLBRGW} aim to extend the set of realizable operations beyond $\SU(2)$).  In fact, $\SU(2)$ operations are
the only accessible with linear optics for qutrits realized with biphotons.

  Basic qutrit spin observables are given by the following three matrices
$$
  S_Z = \pmatrix{1&0&0\cr 0&0&0\cr 0&0&-1\cr},
  \qquad
  S_X = {1\over\sqrt2}\pmatrix{0&1&0\cr 1&0&1\cr 0&1&0\cr},
  \qquad
  S_Y = {1\over\sqrt2}\pmatrix{0&-i&0\cr i&0&-i\cr 0&i&0\cr}.
  \eqdef{spinbasic}
$$
All of them have the three eigenvalues 0 and~$\pm1$.  Respective eigenvectors are given by
$$
  S_Z : \left\{\eqalign{&\ket0,\cr &\ket+,\cr &\ket-,\cr}\right.
  \qquad
  S_X : \left\{\eqalign{&{1\over\sqrt2}\big(\ket+ - \ket-\big),\cr  &{1\over2}\big(\ket+ \pm \sqrt2\ket0 + \ket-\big),\cr}\right.
  \qquad
  S_Y : \left\{\eqalign{&{1\over\sqrt2}\big(\ket+ + \ket-\big),\cr  &{1\over2}\big(\ket+ \pm i\sqrt2\ket0 - \ket-\big).\cr}\right.
$$

\subsection Symmetrisation

Following~\bib{KKLM} we define a linear map from the set of qubit operators to the set of operators over symmetric pairs of qubits.
$$
  \Gamma(U) = {1\over2}(U\otimes I + I\otimes U).
  \eqdef{Gamma}
$$
This map is not multiplicative as $\Gamma(UV)=2\Gamma(U)\Gamma(V)-{1\over2}(U\otimes V+V\otimes U)$ but maps commutators to commutators:
$\Gamma\big([U,V]\big)=2\big[\Gamma(U),\Gamma(V)\big]$.  Operators $\Gamma(U)$ and~$\Gamma(V)$ commute if and only if $U$ and $V$ do.

If $U$ is a $\{\pm1\}$-valued observable.  Then $\Gamma(U)$ can be viewed as a qutrit operator, with eigenvalues~$\pm1$ and~0: if $\ket{v_\varepsilon}$
(with $\varepsilon=\pm1$) are $\varepsilon$-eigenvectors of~$U$, then $\ket{v_\varepsilon}\otimes\ket{v_\eta}$ are ${1\over2}(\varepsilon+\mu)$-eigenvectors
of~$\Gamma(U)$.

\subsection Entanglement

Consider a general (pure) qutrit state
$$
  \alpha\ket++\beta\ket0+\gamma\ket-
  \quad\qbox{with} \abs\alpha^2+\abs\beta^2+\abs\gamma^2=1.
  \eqdef{genqutrit}
$$
Wootters concurrence~\bib{Wootters} for this qutrit is given by
$$
  C = \abs{\beta^2 - 2\alpha\gamma}.
$$
It is known to be a measure of entanglement.  A concurrence equal to 1 means a maximally entangled state, and a zero concurrence means no entanglement.
Entanglement of (spin) qutrits has been discussed in~\bib{BCKS}.

\section The CHSH inequality

  The CHSH~\bib{CHSH} inequality is probably the most simple Bell inequality.  It is based on two parties, two-measurements per party, two
possible issues~$\pm1$ per measurement.  Name Alice and Bob the two parties, and denote $a_1$ and~$a_2$ the values of the two measurements Alice can operate,
and $b_1$ and~$b_2$ the issues of the two measurements Bob can operate.  It is easy to show that, under local realistic assumptions,
$$
  \abs{a_1b_1 + a_1b_2 + a_2b_1 - a_2b_2} \leq 2.
$$
It turns out that quantum systems can violate this inequality to the quantum (Tsirelson) bound $2\sqrt2$~\bib{Tsirelson0}.  It is the case for a quantum
system composed of two qubits: if Alice operates one of two measurements~$A_1$ and~$A_2$ over the left qubit, and Bob operates $B_1$ or~$B_2$ over
the right qubit, the expected value
$$
  \angle{\psi \mid A_1\otimes B_1 + A_1\otimes B_2 + A_2\otimes B_1 - A_2\otimes B_2 \mid \psi}
  \eqdef{E-CHSH}
$$
can reach up the Tsirelson bound~$2\sqrt2$, when for example the four observables are described by the following matrices~:
$$
  \left.
  \eqalign{
    A_1 &: \pmatrix{1&0\cr0&-1\cr},\cr
    A_2 &: \pmatrix{0&1\cr1&0\cr},\cr}
  \right.
  \qquad
  \left.
  \eqalign{
    B_1 &: {A_1+A_2\over\sqrt2} = {1\over\sqrt2}\pmatrix{1&1\cr1&-1\cr},\cr
    B_2 &: {A_1-A_2\over\sqrt2} = {1\over\sqrt2}\pmatrix{1&-1\cr-1&-1\cr},\cr}
  \right.
  \eqdef{AiBj}
$$
and $\ket\phi={1\over\sqrt2}(1,0,0,1)^t$.
As a widely known conclusion this shows that quantum physics cannot be described by local realistic models.

\subsection CHSH for biphotons

  We cannot consider the expected value~\eqref{E-CHSH} when considering biphotons because observables $A_i$ and~$B_j$ are not symmetrical.  However, it turns
out that 
$$
  \angle{\psi\mid \Gamma(A_1)\Gamma(B_1) + \Gamma(A_1)\Gamma(B_2) + \Gamma(A_2)\Gamma(B_1) - \Gamma(A_2)\Gamma(B_2) \mid\psi}
$$
is also upper bounded by~$2\sqrt2$ when applied to biphotons, and this bound is reached for example with the symmetric
state ~${1\over\sqrt2}\big(\ket{00}+\ket{11}\big)$ --- or equivalently with the qutrit ${1\over\sqrt2}\big(\ket{+1}+\ket{-1}\big)$ --- and the $A_i$ and~$B_j$
observables from~\eqref{AiBj}.

  As symmetrical observables, it is possible to express them in qutrit form.  It is easy to check that $\Gamma(A_1)=S_Z$ and~$\Gamma(A_2)=S_X$.  Remark that
$$
  \CHSH(U,V)
  \assign
  U{U+V\over\sqrt2} + U{U-V\over\sqrt2} +  V{U+V\over\sqrt2} - V{U-V\over\sqrt2}
  =
  \sqrt2 (U^2 + V^2).
$$
Moreover, if $W$ is a spin measurement such that $U$, $V$ and~$W$ are pairwise orthogonal, then we know that $U^2+V^2+W^2=2I$ (this is the Casimir invariant).
Thus, for example
$$
  \CHSH(S_Z,S_X)=\sqrt2(S_Z^2+S_X^2)=\sqrt2(2-S_Y^2).
$$
As we yet know, the state~$\ket\psi$ which reach the upper bound~$2\sqrt2$ for $\CHSH(S_Z,S_X)$ is the one such that $\angle{\psi\mid S_Y^2\mid\psi}=0$, 
specifically ${1\over\sqrt2}\big(\ket++\ket-\big)$.  Hence, checking $\angle{\psi\mid S^2\mid\psi}=0$ for some spin measurement, is in fact a check for
maximal entanglement.

\section Spin measurements

\subsection Three basic directions

Consider the general qutrit state~\eqref{genqutrit}.  The probabilities of each issue 0 and~$\pm1$ for each measurement~\eqref{spinbasic} of this state
are respectively
$$
  \eqalign{
     S_Z &: \abs\beta^2,\qquad \abs\alpha^2,\qquad \abs\gamma^2, \cr
     S_X &: {1\over2}\Abs{\alpha - \gamma}^2, \qquad
            {1\over4}\Abs{\alpha \pm \sqrt2\beta + \gamma}^2,\cr
     S_Y &: {1\over2}\Abs{\alpha + \gamma}^2, \qquad
            {1\over4}\Abs{\alpha \mp i\sqrt2\beta - \gamma}^2.\cr}
$$
For random generation, it is suitable that each issue occur with the same probability, for each of the three basic measurements.  The following proposition
explicits the conditions under which this happens.

\label{onethird}
\stat Proposition.  For the four states
$$
  {1\over\sqrt3}\big(\zeta\ket+ \pm \ket0 + \zeta^3\ket-\big)
  \qbox{and}
  {1\over\sqrt3}\big(\zeta^3\ket+ \pm \ket0 + \zeta\ket-\big)
  \qquad\hbox{where $\zeta=\exp(i\pi/4)$}
$$
any of these nine probabilities are~$1/3$.  Moreover they are the only states for which this propery holds.

These four {\sl unbiaised} states are maximally entangled (their concurrence equals~1).
We now choose one of these four states, say
$$
  \ket\psi
  =
  {1\over\sqrt3}\big(\zeta\ket+ + \ket0 + \zeta^3\ket-\big)
  \eqdef{unbiaised}
$$
and we will use it for random generation.

\subsection Spin direction in the plane $x,y$

  We want to express the spin observable, for any direction.  A direction (unit vector) will be described with polar coordinates
$(x,y,z)=(\cos\chi\cos\varphi,\cos\chi\sin\varphi, \sin\chi)$.  For convenience, we will put $c=\cos\chi$, $s=\sin\chi$ and $\theta=\exp(i\varphi)$.

We begin with $\chi=0$.  So we consider spin measurements in the $x,y$-plane with direction defined by the angle~$\varphi$ or more conveniently by the unit
complex number~$\theta=\exp(i\varphi)$.  The spin observable in this direction is
$$
  S_\theta = \cos\varphi S_x + \sin\varphi S_y = {1\over\sqrt2} \pmatrix{0&\theta^*&0 \cr \theta&0&\theta^*\cr 0&\theta&0\cr}
$$
and has eigenvectors
$$
  \left\{\eqalign{
        0:\quad    &{1\over\sqrt2}\big(\theta^*\ket+ - \theta\ket-\big),\cr  
        \pm1:\quad &{1\over2}\big(\theta^*\ket+ \pm \sqrt2\ket0 + \theta\ket-\big).\cr}
  \right.
$$
We can also compute the probabilities of issues 0 and~$\pm1$ when measuring the state~\eqref{genqutrit}.
$$
  {1\over2}\Abs{\theta\alpha - \theta^*\gamma}^2,
  \qquad
  {1\over4}\Abs{\theta\alpha \pm \sqrt2\beta + \theta^*\gamma}^2.
$$

\subsection General spin

For a general direction spin measurement, the spin observable can be written as
$$
  S_{c,\theta}
  = cS_z+sS_\theta 
  = \pmatrix{c&s\theta^*/\sqrt2&0 \cr s\theta/\sqrt2&0&s\theta^*/\sqrt2\cr 0&s\theta/\sqrt2&-c\cr}
$$
and its eigenvectors are given by
$$
  \left\{\eqalign{
      0:\quad    &-s{\theta^*\over\sqrt2}\ket+ + c\ket0 + s{\theta\over\sqrt2}\ket-,\cr 
      \pm1:\quad &{1\over2}\big((1\pm c)\theta^*\ket+ \pm s\sqrt2\ket0 + (1\mp c)\theta\ket-\big).\cr}
  \right.
$$
Note that the 0-eigenvector is maximally entangled (its concurrence equals~1), but the $\pm1$-eigen\-vectors are non-entangled (their concurrence equals~0).
The probabilities of the issues 0 and $\pm1$ when measuring state~\eqref{genqutrit} are given by~:
$$
  {1\over2}\Abs{-s\theta\alpha + \sqrt2c\beta + s\theta^*\gamma}^2,
  \qquad
  {1\over4}\Abs{(1\pm c)\theta\alpha \pm s\sqrt2\beta + (1\mp c)\theta^*\gamma}^2.
$$

\subsection State testing

First observe that the state~$\ket\psi$ given in~\eqref{unbiaised} is the 0-eigenvector of the $S_{c,\theta}$ observable for $c=1/\sqrt3$ and $\theta=\zeta^3$.
Explicitely we have
$$
  S_{1/\sqrt3,\zeta^3}
  =
  {1\over\sqrt3} \pmatrix{1&-\zeta&0\cr -\zeta^*&0&-\zeta\cr 0&-\zeta^*&-1\cr}
  \qqbox{and}
  S_{1/\sqrt3,\zeta^3}^2
  =
  {1\over3} \pmatrix{2&-\zeta&\zeta^2\cr -\zeta^*&2&\zeta\cr \zeta^{2*}&\zeta^*&2\cr}.
$$
Using some measurements with observable $S^2_{1/\sqrt3,\zeta^3}$, one can check that a source of states~$\ket\psi$ does not deviate from~\eqref{unbiaised}
because it is the only state satisfying $\angle{\psi|S^2_{1/\sqrt3,\zeta^3}|\psi}=0$ and because $\angle{\varphi|S_{1/\sqrt3,\zeta^3}^2|\varphi}>0$ for any other
state.

\section Random generation

Here we consider the use of a single qutrit for random generation.  We cannot use non locality properties as in~\bib{PAM+}, but we know that qutrits
exihibits contextuality features, as shown in~\bib{KCBS}.  Random generation using non-contextuality of a qutrit has yet been physically realized, as
reported in~\bib{KJPWF}, but without optimization of the entropy of issues.  Another realization has been reported in~\bib{UZZWSDDK}, where the inequality
KCBS was used to obtain a bound on the generated entropy.  We propose here a different method, and we aim to reach optimality by using the unbiaised
state~\eqref{unbiaised} and suitable measurements.
 
  We need a source of {\bf public} randomness, that is a sequence of random or pseudorandom numbers which can be known by attackers.   This public randomness
can easily be converted in trits (ternary digits) which will be denoted~$(r_1,r_2,\ldots)$.  It will used as input in our generator.  We also need a source
of qutrits $(q_1,q_2,\ldots)$ in the unbiaised state~\eqref{unbiaised}.  For most qutrits~$q_i$ provided by the source, we setup a measurement device in one
of the three basic directions ($S_z$, $S_x$ or $S_y$ if $r_i=0$, 1, or~2 respectively).  The issue of the measurement is also a trit, denoted~$a_i$.  The 
sequence of output trits is a totally uncorrelated to the input sequence, and its security is ensured by the truly random nature of the issues of quantum
measurements.  Moreover, the entropy of the generated sequence is maximal, provided the quantum source delivers qutrits in unbiaised state.

  For this, we need to check the quality of the quantum states source.  This quality is given by the quantum fidelity between the states~$\varphi$ delivered
by the source, and the unbiaised state.  To evaluate it, we (randomly) choose a fraction of the delivered states, and
apply to them a measurement with observable $S^2_{1/\sqrt3,\zeta^3}$ instead.  A statistical analysis of the results of these checks reveals a
defect in the quality of the source if the expected value obtained is above some bound.  In that case, the privacy of the output sequence of
trits is not guaranteed and it should be discarded.  If the expected value of these checks remains below the bound, the output sequence can be used safely
for cryptographic purposes.

  More precisely, the probability that $\angle{\psi|S_{1/\sqrt3,\zeta^3}|\psi}=0$ (or equivalently $\angle{\psi|S^2_{1/\sqrt3,\zeta^3}|\psi}=0$) is the
fidelity of the state~$\ket\psi$ relatively to the 0-eigenstate~$\ket\varphi$ of $S_{1/\sqrt3,\zeta^3}$ given in~\eqref{unbiaised}.   It equals
$$
  F = \abs{\braket\varphi\psi}^2 = {1\over3} \abs{\alpha\zeta^* + \beta - \gamma\zeta}^2.
$$
The following proposition can be used to estimate, within an risk $\delta$ arbitrarilly small, that the fidelity of the source is adequate.

\stat Proposition.  Let $\alpha>0$ and~$\delta>0$.
Let $X_i$ the random variable with takes value~1 when the $i$-th measurement issues 0 and takes value 0 otherwise.  Assume we conduct $\ell$ independent
measurements with $\ell=\ceil{1/4\varepsilon^2\delta}$.  Let $Y={1\over\ell}\sum X_i$.  Then the probability that the random variable $Y$ fails to
estimate~$F$ within $\varepsilon$ precision is less than~$\delta$:
$$
  \prob\{\abs{Y - F} \geq \varepsilon\} \leq \delta.
$$

\proof The $X_i$ are Bernoulli random variables with expected value $F$ and variance~$F(1-F)$.  Then the $Y$ random variable has expected value
$F$ and variance ${1\over\ell}F(1-F)$.  By the Chebyshev inequality we get:
$$
  \prob\{\abs{Y - F} \geq \varepsilon\} \leq {F(1-F) \over \ell\varepsilon^2} \leq {1\over 4\ell\varepsilon^2} \leq \delta.
$$
\qed

\section Conclusion

  We described a quantum number generator, which produces secure random ternary digits sequences with maximal entropy.  The security of this sequence is
ensured by the truly random nature of issues of quantum measurements.  The higher entropy of the output sequence is obtained using unbiaised
states~\eqref{unbiaised}.  The physical realisation of this random generator should be eased by the fact it needs only $\SU(2)$ operations and measurements.

\section Bibliography

\showbib
\bye